**On adjustment for temperature in heatwave epidemiology: a new method and toward clarification of methods to estimate health effects of heatwaves**

Honghyok Kim[1], Michelle L. Bell[2]

[1]Division of Environmental and Occupational Health Sciences, School of Public Health,

University of Illinois Chicago, Chicago, Illinois, USA

[2]School of the Environment, Yale University, New Haven, CT, USA

Corresponding author: Honghyok Kim

Email: honghyok@uic.edu

Telephone: (312) 996-4952.

Address: 1086 SPHPI MC 923, 1603 W. Taylor St., Chicago, IL, 60612 USA

**Conflicts of Interests**

The authors declare they have nothing to disclose.

**Acknowledgement**

This publication was developed under Assistance Agreement No. RD835871 awarded by the U.S. EPA to Yale University. It has not been formally reviewed by the U.S. EPA. Research reported in this publication was also supported by the National Institute on Minority Health and Health Disparities of the National Institutes of Health under Award Number R01MD012769. This work was supported by the Wellcome Trust [216087/Z/19/Z]. Dr Honghyok Kim was further supported by Basic Science Research Program through the National Research Foundation of Korea (NRF) funded by the Ministry of Education (2021R1A6A3A14039711).

**Data availability statement**

We developed an *R* package, *HEAT* (*Heatwave effect Estimation via Adjustment for Temperature*) and made it publicly available at the first author's GitHub (https://github.com/HonghyokKim/HEAT). Researchers can use this package to apply our novel modeling approaches to estimate *HW-Y* relation. Readers can replicate results with *R* code and the dataset.

**Abbreviations**

EHT: Extremely High Temperature

HD: heatwave day

HT: High Temperature

HW: Heatwave

IPCC: Intergovernmental Panel on Climate Change

LT: Low Temperature

MHT: Moderately High Temperature

NHW: Non-Heatwave

**ABSTRACT**

Defining the effect of exposure of interest and selecting an appropriate estimation method are prerequisite for causal inference. Understanding the ways in which association between heatwaves (i.e., consecutive days of extreme high temperature) and an outcome depends on whether adjustment was made for temperature and how such adjustment was conducted, is limited. This paper aims to investigate this dependency, demonstrate that temperature is a confounder in heatwave-outcome associations, and introduce a new modeling approach to estimate a new heatwave-outcome relation: $E[R(Y)|HW=1, Z]/E[R(Y)|T=OT, Z]$, where *HW* is a daily binary variable to indicate the presence of a heatwave; *R(Y)* is the risk of an outcome, *Y*; *T* is a temperature variable; *OT* is optimal temperature; and *Z* is a set of confounders including typical confounders but also some types of *T* as a confounder. We recommend characterization of heatwave-outcome relations and careful selection of modeling approaches to understand the impacts of heatwaves under climate change. We demonstrate our approach using real-world data for Seoul, which suggests that the total effect of heatwaves may be larger than what may be inferred from the extant literature. An *R* package, HEAT (*Heatwave effect Estimation via Adjustment for Temperature*), was developed and made publicly available.

**INTRODUCTION**

Heatwaves have become more frequent, severe, and prolonged due to climate change (1). Timely epidemiological evidence can provide valuable information on how to address public health burdens (e.g., improved heat responses and preparedness) (2-7).

Time-series and case-crossover studies in temperature epidemiology typically investigate single days of high temperature (8, 9), whereas studies on heatwaves focus on consecutive days with extreme high temperatures (10, 11). Understanding the differences between the health impacts of heatwaves and high temperatures is important to inform policies aimed to protect public health, such as heat action plans.

Some studies on heatwaves and health adjust for temperature (12-15) whereas others do not (16, 17). The meaning of heatwave-outcome associations, depending on whether such adjustment was made, is not well studied. Some studies discussed this issue from a statistical standpoint (13, 18). However, understanding the meaning requires conceptualizing the effect of heatwaves and identifying estimation methods relevant to the conceptualized effects (19, 20).

We demonstrate that heatwave-outcome associations depend on whether adjustment for temperature is conducted and how such adjustment is made. We propose a novel modeling approach to estimate heatwave-outcome relations based on our conceptualization of the effect of heatwaves. We demonstrate how modeling approaches impact causal inference using an example with real-world data.

**DEFINING A HEATWAVE**

Heatwaves are not defined consistently in research and policy (11, 21, 22), although they are generally defined based on duration and intensity. For example, the United States National Weather Service defines a heatwave as "a period of abnormally hot weather generally lasting ≥2 days"(22). Heat warning systems sometimes rely on heat indices (23, 24) which are designed to measure thermal comfort by integrating ambient temperature ($T$) and humidity. It is unclear whether heat indices are better causal factors of health outcomes than $T$ alone (25, 26). A multi-country epidemiological study (26) found that humidity does not appear to be a causal factor to non-accidental mortality risk. Numerous epidemiological studies have defined a heatwave based on $T$, without consideration of humidity (11). We follow this convention.

Commonly used methods to define heatwaves considers duration (at least a certain number of continuous days, often ≥2 days) and intensity ($T$ exceeding a specified threshold value, $T_{HW}$ such as the 98$^{th}$ percentile of $T$ distribution)(11). Research often uses a daily binary variable to indicate the presence of a heatwave for day $t$, $HW_t$. For example, using a definition of a heatwave as ≥2 consecutive days with $T \geq T_{HW}$, if $T_t < T_{HW}$, $HW_t = 0$. If $T_t \geq T_{HW}$, and either of the surrounding days exceed $T_{HW}$ (i.e., $T_{t-1} \geq T_{HW}$ or $T_{t+1} \geq T_{HW}$) thereby generating two consecutive days exceeding $T_{HW}$, $HW_t = 1$ meaning that day $t$ is categorized as a heatwave day; and $HW_{t-1}=1$ or $HW_{t+1}=1$. Heat may occur for conditions other than a heatwave with a single day of heat or consecutive days of heat that do not exceed $T_{HW}$. For clarification, we define high temperature, $HT$, as $T$>optimal temperature ($OT$), a temperature at which risk of a health outcome, $Y$, $R(Y)$, is minimal such as minimum mortality temperature (27, 28). We refer to conditions where $HT < T_{HW}$ or $HT \geq T_{HW}$ as moderately $HT$ ($MHT$) or extremely $HT$ ($EHT$), respectively.

We define a heatwave as having ≥2 consecutive days with $EHT$ as has been applied previously (11). There is no standard value for $T_{HW}$, and a variety of $T_{HW}$ approaches have been applied (e.g., 95th, 97th, 98th, and 99th temperature percentiles).

**DEFINING THE EFFECT OF A HEATWAVE**

Defining the effect of exposure of interest is a prerequisite for causal inference (19, 20). Reflecting available public health or surveillance data (10), studies investigated associations of heatwaves with mortality, hospitalizations, and emergency department visits, often with different methods. The findings are sometimes discussed without articulation of how different modeling approaches can affect the meaning of associations (not estimates). Here, we conceptualize the effect of a heatwave and then investigate dependence of heatwave-outcome associations on estimation methods.

*Multiple causal pathways for the effect of heatwaves*

There exist many potential causal pathways for the effect of heatwaves, such as difference in susceptibility between biological stress by cumulative exposure to $EHT$ and the stress by instant exposure to $EHT$ or $MHT$, adaptive behaviors to extreme heat, and disrupted social, economic, and health systems during long-lasting $EHT$ (7, 29-33). The Intergovernmental Panel on Climate Change (IPCC)'s Sixth Assessment Report (1) highlights cascading effects by weather-related disasters, meaning that heatwaves can indirectly increase health risks by impacting socioeconomic infrastructures in addition to direct health impacts (i.e. physiological effects by heat).

Figure 1A is the simplest directed acyclic graph (DAG) that shows the causal effect of *HW* on a health outcome (*Y*): *HW→Y*. *HW* indicates the occurrence of a heatwave in ambient environment, not exposure to a heatwave. Figure 1B is an extension of Figure 1A including exposure to a heatwave (*E*): *HW→E→Y*. Figure 1C is an extension of Figure 1B that considers protection (*P*) to avoid *E*. For example, people can use air conditioning, visit cooling centers, or avoid outdoor activities during a heatwave: $HW→P_1→E→Y$. There may be a set of factors, *Q*, that impact $P_1$. For example, some people may need to stay outside (e.g., workers of essential services) or may not avoid exposure (e.g., not using air conditioning due to worry about energy costs (5); limited personal protection equipment for outdoor workers). There is another protection regardless of *HW* such as indoor built-environment (e.g., architecture designed with natural ventilation and cooling): $P_2→E→Y$. *HW* is unlikely to be perfectly correlated with *E*.

Figure 1D is another extension of Figure 1B, including the effect of *HW* on the infrastructure (*I*) (*HW→E→I* and *HW→I*) and directly heat-related outcome, $Y_1$ and other outcome, $Y_2$ due to impacted *I* such as cascading effects (1): $I→Y_1$ and $I→Y_2$. *I* can collectively denote socioeconomic infrastructure (e.g., health systems, transportation networks, power supplies). For example, whether health systems are heat-resilient to prevent a systemic failure of health services should be considered (7). Critical patients with heat-related illness can be at risk of losing receiving timely treatment if emergency departments are overwhelmed. Overburdened hospitals may struggle to provide timely medical treatments to patients with other conditions, exacerbating their health issues, as highlighted in $Y_2$. A heatwave can increase the risk of power outages (34) and traffic chaos (e.g., increased risk of traffic crashes (35), road closures, rutting, and buckling (36)) may occur, impacting accessibility to necessary services and hospitals. Grocery stores and pharmacies may temporarily close, especially under national/state of emergency and/or power outages, resulting in water, food, and medicine insecurities. Figure 1E, a cyclic graph, integrates Figures 1C–D, where the compromised infrastructure can impact $P_1$, such as power outages resulting in the failure of air conditioning: $HW \rightarrow I \rightarrow P_1$.

We define the relation between $HW$ and $Y$ ($Y_1+Y_2$) as the total effect of $HW$, representing the population-averaged effect, which is the main interest here. We refer to the relation between $HW$ and $Y_1$ through $E$ and $P$ as a direct effect; between $HW$ and $Y_1$ or $Y_2$ through $I$ as indirect effects. These definitions presume that physiological pathways (e.g., heat toxicity through heat exposure) are subsumed in the direct effect.

*Isolating the effect of heatwaves from the effect of temperature*

The health effect of $HW$ is policy-relevant, and the effect may differ from a single day of $MHT$ or $EHT$ while they may share some causal pathways. Health systems, transportation networks, and power grids are more likely to be disrupted during heatwaves. Many anecdotes and news reports demonstrate power outages and overburdened emergency departments for consecutive days of extreme heat. Heat safety actions and responses can be activated if temperature exceeds a certain intensity threshold, including disaster preparedness of health systems (7), governmental subsidies for electricity bills (5), electricity demand management (37), and railroad speed limits (38), which further substantiates the effect of $HW$ and $EHT$. Evidence shows higher association with fatal traffic crashes for extreme heat than for moderate heat (35) and for road buckling (36).

Figure 2A shows DAGs for the effect of *T* on *Y*. We propose that temperature effect should be regarded as two parts: *HW* and temperature not related to a heatwave ($T^{NHW}$) (Figure 2B). Some investigators may be interested in *T-Y* relation including extreme heat (waves) (Figure 3: *T* effect). We focus on *HW-Y* relation (Figure 3: *HW* effect)

*Epidemiological definition of HW-Y relation*

We define *HW-Y* relation as $E[R(Y)|HW=1, Z]/E[R(Y)|T=OT, Z]$, where *Z* is a set of confounders. *Z* includes typical confounders but also confounding by $T^{NHW}$ and confounding by lagged *HW*. The latter two have not been widely acknowledged in heatwave epidemiology, which will be discussed in the next section.

*Lag effects and cumulative exposure effects on R(Y) of heatwave days*

Researchers often use terms, lag(ged) effects and cumulative exposure effects, while these terms are sometimes not precisely defined. *HW* is defined using a lag-structure of *T* such that the effect of *HW* is related to lagged effects and cumulative exposure effects of *T*.

In disciplines outside environmental epidemiology, the term *lag* may have the connotation of *induction period* in epidemiology taxonomy (39). We adopt the term *induction period* for clarification, henceforth. *lag(ged) effect* should mean the effect of exposure with *an induction period*. For example, there exist Lag0 effect (i.e., the induction period is zero day, thus instantaneous effect) of one-day exposure to *HT*, Lag1 effect (i.e., the induction period is one day) of one-day exposure to *HT*, and so on. We call them L0(1d), L1(1d), and so on. We relegate examples about lagged effects to Appendix A. Effects of cumulative exposure to *HT* are possible. For example, one-day exposure may not be enough to exceed a certain biological threshold to manifest the effect of *HT* (e.g., thrombosis advanced by one-day heat exposure may not be adequate to progress pathological responses leading to death for some people). We refer to Lag0 effect of two-day exposure as L0(2d), refer to Lag1 effect of two-day exposure as L1(2d), refer to Lag0 effect of three-day exposure as L0(3d), and so on.

Figure 4 illustrates how lag and cumulative exposure effects constitute $R_t(Y)$ for non-heatwave days (non-HDs) (*u-2*, *u-1*) and heatwave days (HDs) (*u*, *u+1*). The size of increased risks and *T* time-series are hypothetical for illustration. The effect of *HW* consists of the increased risks by *EHT* of HDs ($EHT^{HW}$) (red boxes). *R(Y)* on HDs can also be increased by *EHT or MHT* of non-HDs ($EHT^{NHW}$ or $MHT^{NHW}$) via lag effects (e.g., L1, L2) and/or *partial* cumulative exposure effects (e.g., L0(2d), L1(2d), L0(3d)) (orange boxes), which is confounding by $T^{NHW}$ (See the next section for detail).

*Counterfactuality*

Many counterfactuals for *HW-Y* relation could exist: the concept that a heatwave increases *R(Y)* from 'a' reference risk (i.e., the risk would have been if the episode had not occurred). This counterfactual statement can reflect many forms. The defined *HW-Y* relation is based on E[*R(Y)*|*T=OT*, *Z*], meaning what *R(Y)* would have been if there had been no effects of *T* deviating from *OT*–the absence of the lag effects/cumulative exposure effects of *T* on *R(Y)* at *T=OT*). Many studies have estimated heat-related premature mortality as *HT* deviated from *OT* (8, 9).

A reference risk could also be what *R(Y)* would have been with $T<T_{HW}$. Some studies estimate an excess mortality risk during a heatwave event by comparing the mortality rate during the event with a historically mean mortality rate during non-heatwave periods (40, 41). That periods could include days with $T<OT$, days with $T=OT$, and days with $T<T_{HW}$. Such average risk appears to serve as another counterfactual risk because the average may differ from the mortality risk at *OT*.

**DEPENDENCY OF *HW-Y* ASSOCIATION ON ADJUSTMENT FOR TEMPERATURE**

Traditionally, *HW-Y* association is estimated by comparing *R(Y)* on HDs and *R(Y)* on non-HDs (11). For example, suppose a Poisson regression model in time-series studies,

$$\log(E[Y_t]) = \alpha + \beta_1 HW_t + confounders \text{ (Model 1)}$$

where $Y_t$ is *Y* at day *t* and $\exp(\beta_1)$ is rate ratio (RR), which could be approximately risk ratio when temporal change in the population-at-risk is negligible (e.g., mortality in a general population). Confounders typically include seasonality, time-trend, and humidity.

We revisit two traditional methods to address *T* (12-18) in efforts to disentangle effects of *HW* from those of *HT*. Some studies do not adjust for *T* when estimating effects of *HW* (16, 17) (Model 1) because *EHT* is one component of *HW* (i.e., intensity). We refer this to as Traditional Approach #1. Other studies adjusted for basic *T* (e.g., *T* of the same day (Lag0), previous day(s) (e.g., Lag1, Lag2 temperatures (12-15))). We refer to this as Traditional Approach #2. We use $T_t$, $T_{t-1},...,T_{t-L}$ to denote Lag0, Lag1,…,Lag*L T*.

*HW-Y* association not adjusted for *T* indicates the difference between *R(Y)* on HDs and a weighted average of *R(Y)* on non-HDs. The latter risk would be averaged from that of three different types of non-HDs: non-HDs with $T<OT$, non-HDs with $T=OT$, and non-HDs with $T>OT$, the latter of which is a day with $HT$ but not satisfying the heatwave definition. The weight would be the number of days for each type of non-HDs. Thus, the counterfactual risk of this $HW$-$Y$ association would be $R(Y)$ at an unclear $T<T_{HW}$. However, this traditional approach may not be relevant for causal inference on the effect of heatwaves because $HW$-$Y$ association could be confounded by $T^{NHW}$ due to the backdoor path of $Y$–$T^{NHW}$–$T$–$HW$ (Figure 2B). $T^{NHW}$ is correlated with $HW$: for $m\geq0$, if $HW_{t-m}=0$, $|T^{NHW}_{t-m}|>0$ and if $HW_{t-m}=1$, $T^{NHW}_{t-m}=0$, which is perfect negative correlation. For $m=0$, this confounding corresponds to the aforementioned counterfactual risk because the effect of $T^{NHW}_t$ on $R_t(Y)$ exist for only $HW_t=0$, not $HW_t=1$. There exist correlations of $HW_t$ with $T^{NHW}_{t-l}$ and correlations of $HW_{t-l}$ with $T^{NHW}_t$, which comes from autocorrelation between $T_t$ and $T_{t-l}$ for $l>0$. If there exist lagged effects of (cumulative) exposure of $T^{NHW}$ or $HW$, which is likely (9, 13), this association is further confounded by these effects: backdoor paths of $Y_t$–$T^{NHW}_{t-l}$–$T_{t-l}$–$T_t$–$HW_t$ and/or $Y_t$–$HW_{t-l}$–$T_{t-l}$–$T_t$–$HW_t$ (when lag effects of $HW$ are not considered in the association (42)). This confounding by $T^{NHW}_{t-l}$ and/or by $HW_{t-l}$ can be articulated by clarifying $R(Y)$. $R_t(Y)$ on HDs can be increased by $T^{NHW}_{t-l}$ (orange boxes on $u$ and $u+1$ in Figure 4). In a similar vein, $R_t(Y)$ on non-HDs can be increased by $T^{NHW}_{t-l}$ but also $HW_{t-l}$.

Traditional Approach #2 adjusts for basic $T$. For example, in a piecewise Poisson regression,

$$\log(E[Y_t]) = \alpha + \beta_2 HW_t + \sum_{j_L=0}^{J_L} \gamma_{L,j_L} T^{LT}_{t-j_L} + \sum_{j_H=0}^{J_H} \gamma_{H,j_H} T^{HT}_{t-j_H} + confounders \quad \text{(Model 2)}$$

where $T^{LT}=OT-T$ if $T\leq OT$, which indicates low temperature ($LT$) and $T^H=T-OT$ if $T>OT$, which indicates $HT$, $\gamma_{LT}$ indicates the regression coefficient for $LT$, and $\gamma_H$ indicates the regression coefficient for $HT$. $J_H$ is the maximum lag period for $J_H+1$ of the effective exposure time-window for $HT$. $J_L$ is the maximum lag period for $J_L+1$ of exposure time-window for $LT$. For illustrational simplicity, we adhere to these segmented linear terms, acknowledging that non-linear terms (e.g., distributed lag non-linear models (43)) may be relevant in adjusting for *T*.

*HW-Y* association adjusted for basic *T* does not fully represent the defined *HW-Y* relation because most of the effect of *HW* is explained away by the adjustment. Figures 5A, 5B, and 5C conceptually illustrate the meaning of this adjustment. This adjusted association is called an added effect of heatwaves (18), being interpreted as how much $EHT^{HW}$ additionally increases risk after the risk from increased single days of $T$ ($T_t,…,T_{t-l}$).

We note that this interpretation should require two previously unnoticed assumptions: 1) $EHT^{NHW}$ does not have any lag effects and contribution to cumulative exposure effects of $EHT^{HW}$; and 2) the effect of $EHT^{NHW}$ represented by basic *T*s is homogeneous with the effect of $EHT^{HW}$ represented by those *T*s. If both do not hold, confounding by $T^{NHW}$ may not be fully controlled. For illustration, suppose that addition of $T_t$ to a regression model fully adjusts for $EHT^{NHW}_t$-*Y* relation and $EHT^{HW}_t$-*Y* relation. *HW-Y* association adjusted for $T_t$ should represent $EHT^{HW}_{t-1}$-*Y* relation (and $EHT^{HW}_{t-l}$-*Y* relation for $l \geq 2$ if some heatwaves last for ≥3 days), to ensure that interpretation. If there exist lag effects of $EHT^{NHW}$, these effects would be subsumed into the *HW-Y* association because $EHT^{NHW}$ is correlated with $EHT^{HW}$. And, suppose that adjustment for $T_{t-1}$ is also made. If $EHT^{NHW}_{t-1}$-*Y* relation is heterogenous with $EHT^{HW}_{t-1}$-*Y* relation, the *HW-Y* association may capture the remaining of $EHT^{NHW}_{t-1}$-*Y* relation to some degree, that may not be fully captured by the coefficient of $T_{t-1}$ because Model 2 does not separately specify these two.

To control confounding by $T^{NHW}$, adjustment for a lag-structure of $T$ needs to be adequate (Figures 5A and 5B). The adjusted *HW-Y* association would go toward the null if more lagged variables of $T$ are adequately controlled (Figure 5C). The null adjusted *HW-Y* association is sometimes interpreted as the absence of the added effect of heatwaves, which does not mean the absence of the effect of *HW*. The meaning of the null association would be that, besides the risk increase by $EHT^{NHW}$ and $EHT^{HW}$ being already subsumed into *T-Y* association, there is no additional increase in *R(Y)* by $EHT^{HW}$ (and by $EHT^{NHW}$ if the first assumption does not hold). Another issues are that the degree of the adjustment depends on how to model $T$ and the counterfactual risk depends on it. We relegate technical details to Appendix B.

Thus, neither of the traditional approaches can fully estimate the defined *HW-Y* relation. Therefore, we propose a new method to estimate the defined *HW-Y* relation by selectively adjusting for $T$ that contributes to confounding (Figure 5D). We replace $T^{HT}$ in Model 2 with $T^{HT*}$. For time *t*, we set $T_{t-l}^{HT*}=T_{t-l}^{HT}$ if $HW_t=0$ and $T_{t-l}^{HT*}=0$ if $HW_t=1$ ($l≥0$), which adjusts for only the effects of $HT^{NHW}$ and $EHT^{HW}$ on *R(Y)* of non-HDs and the effects of $EHT^{HW}$ on *R(Y)* of non-HDs following the HDs. This specification does not control the effect of $HT^{NHW}$ on *R(Y)* of HDs. So, we introduce a vector, *V,* to adjust for only it. If day *t* is the first day of a heatwave, $V_t = (T_{t-1}^{HT}, T_{t-2}^{HT}, ..., T_{t-K}^{HT})$, which adjusts for the effects of $HT^{NHW}$ on *R(Y)* of the first day of a heatwave (see *u* in Figure 5D). If *t* is the second day of a heatwave, $V_t = (0, T_{t-2}^{HT}, ..., T_{t-K}^{HT})$. Note that the first component is zero, not to control the effect of $EHT^{HW}$ of the first day of *HW* on *R(Y)* of the second day of *HW* (see *u+1* in Figure 5D). If *t* is the $K^{th}$ day of a heatwave, $V_t = (0,0, ..., 0, T_{t-K}^{HT})$. *K* can be the maximum duration of heatwaves in dataset minus 1 or $J_H$. So, the model becomes:

$$\log(E[Y_t]) = \alpha + \beta_3 HW_t + \sum_{j_L=0}^{J_L} \gamma_{L,j_L} T_{t-j_L}^{LT} + \sum_{j_H=0}^{K} \gamma_{H,j_H} T_{t-j_H}^{HT*} + \delta V_t + confounders \text{ (Model 3)}.$$

$\delta$ is a vector of the coefficients of each element of $V_t$. We provide our *R* package to help researchers apply this novel approach (Appendix C). For estimation of a lag effect of *HW* (instead of adjusting for this effect by $T_{t-l}^{HT*}$) and how to adjust for temperature using spline, please see Appendix C.

**APPLICATION**

*Methods*

We conducted time-series analysis for Seoul, South Korea, 2006–2013 to demonstrate how *HW-Y* association can vary by modeling approaches to address temperature. This dataset, which has been described previously (44), includes daily 24-hour average of temperature and $PM_{10}$, daily 8-hour moving average maximum of $O_3$, holidays, calendar time, and daily non-accidental deaths (International Classification of Disease 10th Revision, A00-R99).

Only summer seasons (June–August) were analyzed. *T* was daily 24-hour mean temperature. We defined heatwave as *T* exceeding the 99$^{th}$ percentile (28.7°C) of the year-round *T* distribution for at least two consecutive days (11). We fit quasi-Poisson regression models and Models 1–3. We conducted analyses with NCS of $T_t$ and NCS of $T_{t-1}$ without $HW_t$ to identify *OT* (21.7°C). Adjustment for only $T_t$ and $T_{t-1}$ was considered because we found no association of longer lags with mortality using natural cubic splines (NCS). We used piecewise Poisson regressions because there was no difference in association of *T* with mortality between the piecewise regression and regression with NCS of *T*. NCS of the day of season, dummy variables of year, and interaction between these two were included to adjust for time-trend and seasonality.

Dummy variables of day of the week and NCS of two-day moving average of $PM_{10}$ and $O_3$ were added. We also conducted simulation analyses to replicate findings (See online supplementary material)

*Results*

For Models 1 and 2, *HWs* were associated with a -3.2% (95% confidence interval [95% CI]: -10.6, 4.9) and a -3.5% (95% CI: -10.8, 4.5) change in non-accidental mortality, respectively. Alternatively, for Model 3, the association was a 15.2% (95% CI: 1.3, 30.9) change in mortality. The association between *HW* and non-accidental mortality for the sum of Lag0 and Lag1 effects of *HW* was 20.8% (95% CI: -2.2, 49.3). Also, for Model 3, adjustment for temperature using splines showed estimates of 14.7% (95% CI: 0.9, 30.4) for Lag0 effect of *HW* and 19.3% (95% CI: -3.2, 47.0) for Lag0 and Lag1 effects of *HW*, consistent with the original estimates. Simulation analyses replicated these findings (See online supplementary material).

**DISCUSSION**

Clarifying the effect of interest and using appropriate epidemiological methods relevant to estimate the effect is essential for causal inference. Our conceptualization for the effect of heatwaves considers multiple causal pathways, being aligned with IPCC's recent report on impacts of climate change, with cascading effects, and is policy-relevant. We proposed novel modeling approaches to estimate *HW-Y* relation by addressing temperature adjustment. Our data analysis indicates that the total effect of heatwaves on mortality in Seoul, South Korea is higher, based on our novel approach, than what would have been estimated using traditional approaches, suggesting that the traditional approaches may underestimate the full health impact of heatwaves.

Epidemiological research on separating direct and indirect effects will provide unique insights into heat responses and preparedness. This requires additional data that are rarely used in studies of heatwave epidemiology, as commonly used datasets, including our analyses presented for Seoul, addressed only total effect. For example, to identify indirect effects of *HW* via *I*-related pathways, data for compromised infrastructure, such as overburdened health systems and power outages, are necessary. Some studies used data on the prevalence of air conditioning that can indirectly imply *P*, meaning that the prevalence may be a surrogate of *HW-E* correlation and *HW-Y* relation may differ across strata of the prevalence. *P* or *I* may be seen as an effect modifier or a mediator, depending on how it is conceptualized, and modeling approaches and interpretations of results may differ accordingly. To separate $Y_1$ and $Y_2$, outcome classification should be performed considering causal pathways in future research.

The use of *HW-Y* relation instead of *E-Y* relation has many advantages. *HW-Y* relation represents the health impacts of heatwaves through comprehensive causal pathways, which is policy-relevant. Government agencies and community organizations have policies related to the occurrence of heatwaves in their jurisdictions (e.g., heatwave action plans) and are interested in how public health is impacted by heatwaves in the population. *HW-Y* relation would be specific to local contexts because of possibly heterogenous pathways, reducing generalizability and transportability (45). This highlights why location-specific *HW-Y* relation is needed for public health protections. While meta-analyses suggest consistently positive estimates of *HW-Y* associations (e.g., *HW*-mortality association), estimates vary across study populations (10, 11). The use of *HW* instead of the use of *E* does not necessarily mean that exposure misclassification arises (Appendix D).

While no standard definition for a heatwave exists, we followed the convention of defining a heatwave based on intensity and duration (11). Stakeholders often use weather-based metrics and health impact-based approach (11, 21). The implications of how different definitions of a heatwave impacts health estimates merits further research. We suggest that a health impact-based approach should consider total, direct, and indirect effects and policy relevance and clearly define counterfactual risk.

How to adjust for temperature should consider assessments on *T-Y* association regarding its form (e.g., *T-Y* association curve and a relevant lag period). *T-Y* association may differ by seasons and study populations, so that researchers should evaluate *T-Y* association upfront. More flexible regression models (e.g., splines and a longer lag period of *T*) may also be applied (See Appendix C.3).

We note that our results from the Traditional Approach #1 differ from findings in previous studies that indicated statistically significant positive associations between heatwaves and mortality in South Korea (16, 47). Notably, there are several differences: the study period and study population (multi-city vs. single city), how adjustment for time trends and seasonality was performed, and whether adjustment for air pollutants was performed. One study found positive associations but those estimates were not statistical significance (15). These issues merit additional investigations.

In conclusion, causal inference on health impacts of heatwaves should be based on clarifications of the effect of interest and estimation methods relevant to a defined effect. Diligent selection of modeling approaches and careful interpretation of heatwave-outcome associations are necessary to understand actual impacts of heatwaves under climate change and guide policymakers. Our findings indicate that traditional modeling approaches may underestimate the public health burden of heatwaves.

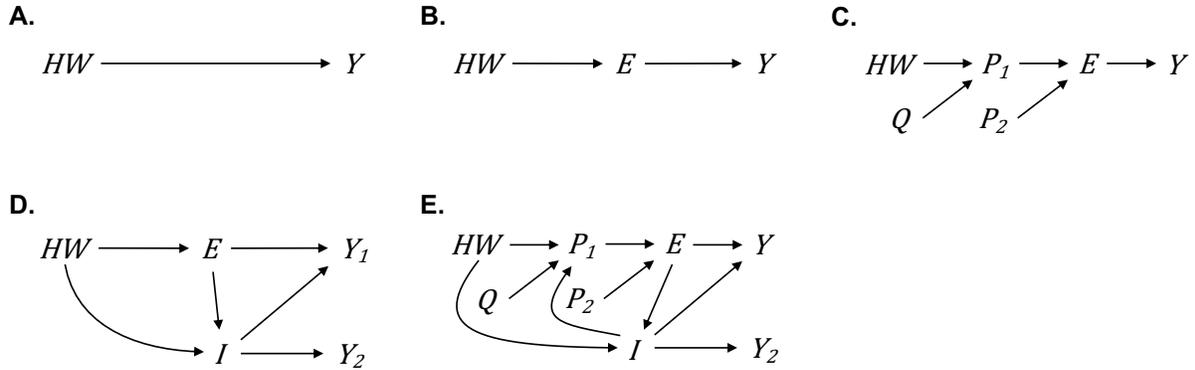

**Figure 1. Directed acyclic graphs (A–D) and a directed cyclic graph (E) for clarifying the effect of heatwaves.**

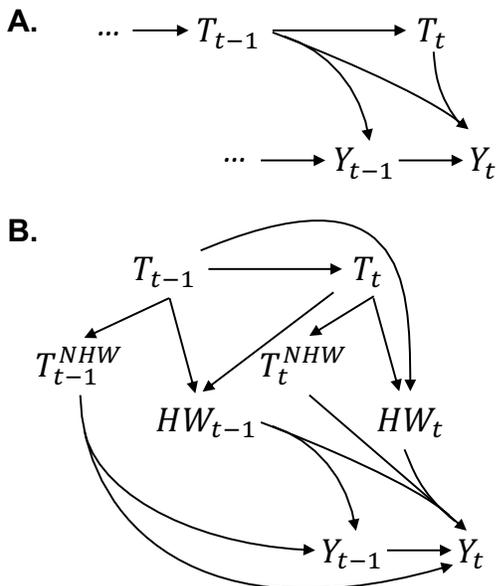

**Figure 2. Directed acyclic graphs for the relation between ambient temperature ($T$) and a health outcome ($Y$). A. A simple conceptualization; B. A conceptualization of distinguishing a heatwave ($HW$) and non-heatwave related $T$ ($T^{NHW}$).**

Note. Subscripts $t$, and $t$-$1$ denote time $t$ and time $t$-$1$. $T_{t-1} \rightarrow T_t$ and $Y_{t-1} \rightarrow Y_t$ denote an autocorrelation process.

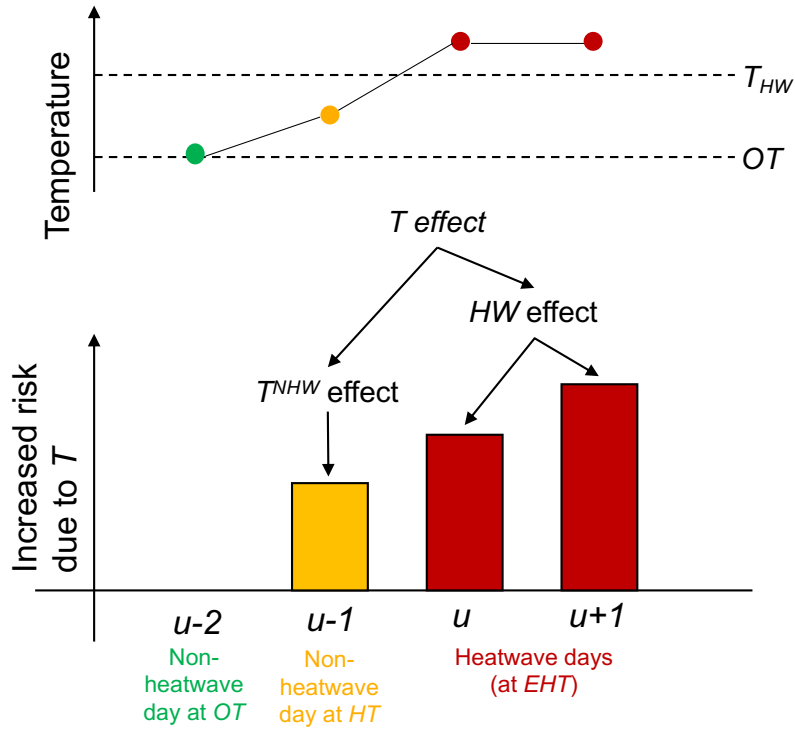

**Figure 3. An illustration of the difference between the effect of *HW* and the effect of $T^{NHW}$.**

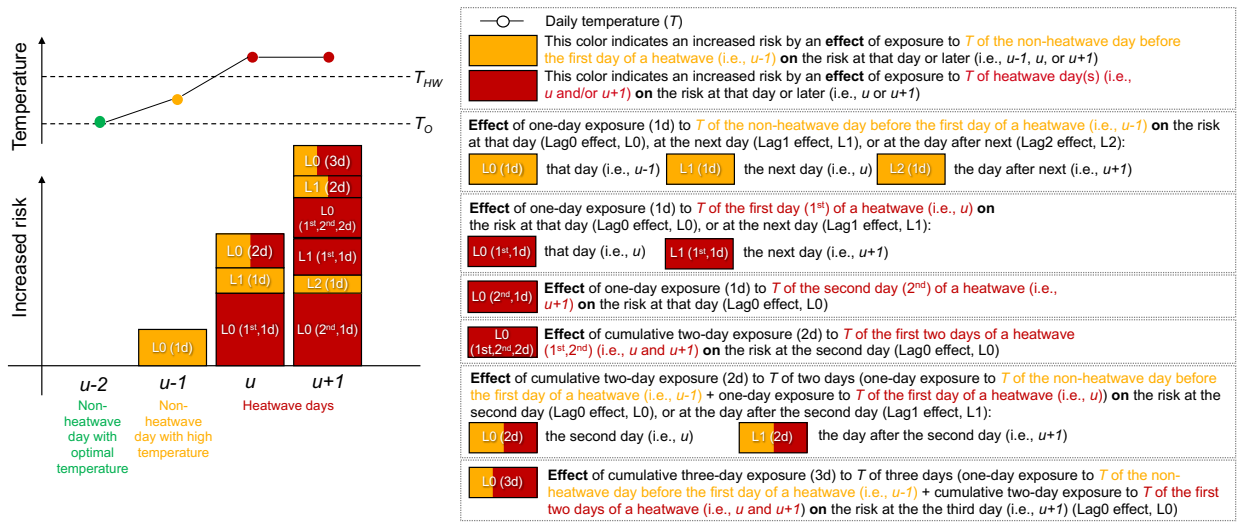

**Figure 4. An illustration of how the increased risk on heatwave days and non-heatwave day can be decomposed into different increased risks by lag effects and cumulative exposure effects of *T*.**

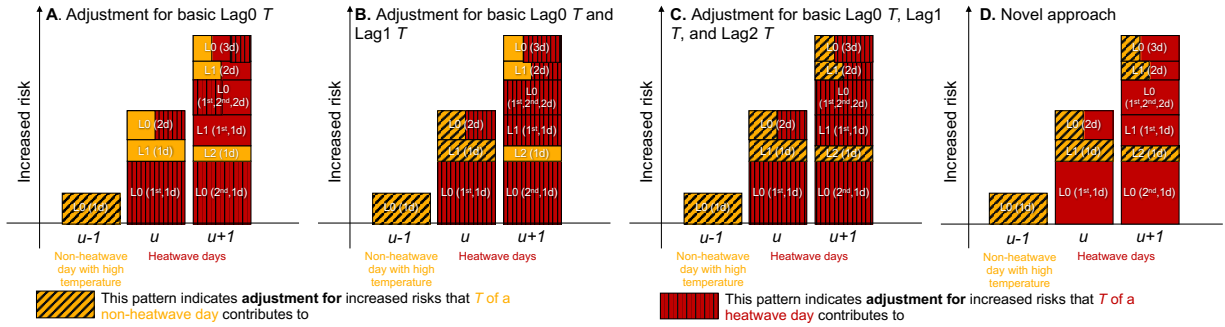

**Figure 5. An illustration of how adjustment for temperature can differently adjust for increased risks.**

**Appendix A. Examples of lag effects and cumulative exposure effects**

For mortality and hospitalizations, a few days of induction periods are plausible. There are other pathways aside from biological pathways. For example, suppose that a person was exposed to heat, had symptoms, and got hospitalized. That person received a medical treatment in an intensive care unit for two days but was pronounced dead unfortunately. In this case, the induction period for the effect of exposure to heat on this death is two days. Suppose that a person with diabetes was unable to use insulin treatments because his/her refrigerator was out of electricity due to power outage during a heatwave episode. That person got hospitalized a few days later when hospitals that were overburdened during the episode became available but eventually passed away due to cardiovascular complications. In this case, the induction period for the (indirect) effect of that heatwave on the death or the hospitalization is not zero day. Suppose two people who lived in a different community experienced cardiac arrest during a heatwave. One person was timely transported to a hospital by an ambulance and survived. Transportation of the other person was delayed due to traffic chaos and his/her condition was exacerbated. He/she passed away even after a few days of medical treatments. The induction period for the (total) effect of that heatwave on the death of the second person is a few days. The findings of many epidemiological studies suggest acute effects of *HT* on mortality (e.g., L0(1d), L0(2d), L1(1d), L1(2d))(9, 14, 27, 48-50), which are plausible, considering multiple causal pathways including biological plausibility and cascading effects (Figure 1).

**Appendix B. Added effect of heatwaves from Traditional Approach #2**

Figure B1A illustrates statistical concepts of Traditional Approach #2. Basic *T* is expected to capture the increased risks by *T* deviated from *OT*. *HW* is expected to capture the increased risk that is not fully explained by basic *T*.

Note that splines may be used instead for basic *T*. In this case, the increased risk not fully explained by basic *T* may differ because a spline function may explain some *R(Y)* at $EHT^{HW}$ (Figure B1B). The degree to which basic *T* explains increased risks depends on the specification of a function of *T* and distribution of data points, meaning that the estimand of Traditional Approach #2 differs because the reference risk (i.e., the bottom dotted lines in Figure B1) differs.

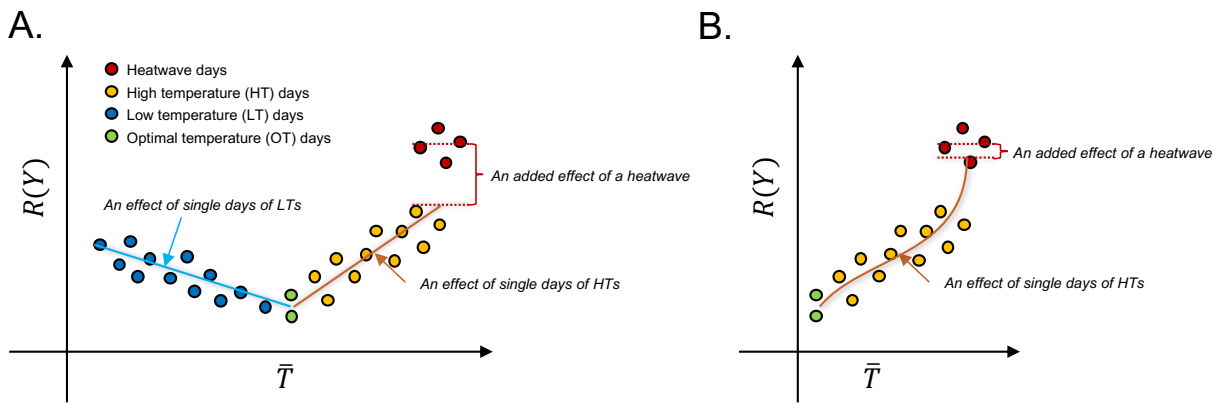

**Figure B1. An illustration for the statistical concept of Traditional Approach #2. Each circle indicates a hypothetical data point. $\bar{T}$ indicates an (weighted) average of temperature lagged variables, based on the link between weights of a moving average and coefficients of distributed lag (non-linear) models (51).**

**Appendix C. The novel modeling approaches.**

*R* package, *HEAT*

We developed an *R* package, *HEAT* (*Heatwave effect Estimation via Adjustment for Temperature*) and made it publicly available at the first author's GitHub (hyperlink to be added after the blinding review). Researchers can use this package to apply our novel modeling approaches to estimate *HW-Y* relation. Readers can replicate results with *R* code and the dataset.

*C.1. Additional illustrations for Model 3*

This sub-section provides additional illustrations for Model 3. We illustrate this with an excerpt of the time-series dataset for Seoul, South Korea (Figure C1). All variables except for $FHW_{t-1}$, $T_{t-1}^{HT***}$, $T^*$s, and $T^{**}$s were introduced in the main text; we will introduce these below. For illustrational simplicity, we consider only Lag0 $T$ ($T_t$) and Lag1 $T$ ($T_{t-1}$). Our *R* package, *HEAT*, allows for ≤Lag7 in analysis. Recall that we need to adjust for confounding by $T^{NHW}$ and lagged *HW*. Simultaneously we must not adjust for increased risks by $EHT^{HW}$ on HDs (Figure 5).

We introduce the variables, $T_t^{HT*}$ and $T_{t-1}^{HT*}$ that were introduced in the main text. $T_t^{HT*}$ and $T_{t-1}^{HT*}$ adjust for the increased risks by $HT^{NHW}$ on non-HDs because their values are >0 when *HW*=0 (See the red boxes for $T_t^{HT*}$ and $T_{t-1}^{HT*}$ in Figure C1B). Unlike $T_t^{HT}$ and $T_{t-1}^{HT}$, they do not adjust for the increased risks by $EHT^{HW}$ on HDs because their values are zero when *HW*=1 (See the yellow-shaded rows with the green boxes in Figure C1B). $T_{t-1}^{HT*}$ also adjusts for one-day lag effect of *HW* on the following non-heatwave day (See the purple boxes in Figure C1B), which can be separately estimated instead of being controlled by $T_{t-1}^{HT*}$ (See the next subsection). Model 3 also adjusts for $T_t^{LT}$ and $T_{t-1}^{LT}$ because there may be increased risks by *LT* (See the red boxes for $T_t^{LT}$ and $T_{t-1}^{LT}$ in Figure C1B). By adjusting for all these, the counterfactual risk of *HW-Y* relation is set to be the risk at *OT*. This is because the increased risks on non-HDs by *T* deviating from *OT* (i.e., *LT*, *HT$^{NHW}$*, and *EHT$^{HW}$*) and by *EHT$^{HW}$* are controlled.

Lastly, we introduce the variable, $V_t$, that was introduced in the main text. This variable adjusts for the increased risks by *HT$^{NHW}$* on HDs (See the value in the blue box for $V_t$ in Figure C1B). This adjustment is necessary because lag effects or any partial cumulative exposure effects of *HT$^{NHW}$* are not the effect of heatwaves (e.g., orange boxes on days *u* and *u+1* in Figures 4 and 5).

*C.2. Lag effects of HW*

This sub-section illustrates how to separately estimate lag effects of *HW* instead of adjusting for them by $T_{t-1}^{HT*}, T_{t-2}^{HT*}, \ldots$, and so on. We start with the values of $T_{t-1}^{HT*}$ in the red boxes in Figure C1C. On the non-HD following the end of a heatwave, this variable indicates *T* of the end of this heatwave minus *OT*, representing the one-day lagged *EHT$^{HW}$* of the end of this heatwave. If the estimand of interest does not include any lag effects of *HW*, they must be considered as a confounder because they impact *R(Y)* on non-HDs. They must be controlled. We introduced $T_{t-1}^{HT*}$, $T_{t-2}^{HT*}, \ldots, T_{t-L}^{HT*}$ earlier and they can adjust for the lagged effect. If the estimand of interest includes lagged effects of *HW*, they must not be controlled but estimated separately, which is the focus of this sub-section.

To estimate lagged effects of *HW*, we introduce new variables. For example, *FHW$_{t-1}$* is a one-day lagged variable for *HW$_t$* but note that this is an unique lagged variable of *HW$_t$*: *FHW$_{t-1}$*=1 only if *t-1* is the first day after a heatwave; otherwise this is zero. (See the blue boxes for *FHW$_{t-1}$*, in Figure

C1C). And we must not adjust for lagged effects of *HW* by using $T_{t-1}^{HT*}, T_{t-2}^{HT*}, ..., T_{t-L}^{HT*}$. We introduce $T_{t-1}^{HT**}, T_{t-2}^{HT**}, ..., T_{t-L}^{HT**}$ instead. $T_{t-l}^{HT**}$ is the *l*-day lagged variable of $T_t^{HT*}$. For example, note that the zero value of $T_{t-1}^{H**}$ in the green boxes in Figure C1C, which differs from $T_{t-1}^{H*}$. How to generate lagged variables makes difference. One way is $T_{t-l}^{HT*}=T_{t-l}^{HT}$ if $HW_t=0$ and $T_{t-l}^{HT*}=0$ if $HW_t=1$ for $l>0$. This way creates $T_{t-l}^{HT*}$. Another is to generate lag variables using $T_t^{HT*}$ itself. This way creates $T_{t-l}^{HT**}$.

We note another difference between $T_{t-l}^{HT*}$ and $T_{t-l}^{HT**}$. $T_{t-l}^{HT**}$ contains the information of $V_t$ (See the values in the blue boxes for $T_{t-1}^{HT**}$ and $V_t$ in Figure C1C). So, if $T_{t-l}^{HT**}$ is used, $V_t$ may be redundant. This information will be used to adjust for effects of *HT^NHW* on HDs. One may want to use $T_{t-l}^{HT***}$ that does not contain the information of $V_t$: $T_{t-l}^{HT***}=T_{t-l}^{HT**} - $ *l*-th component of $V_t$ for $l \geq 1$ only if *t* is a HD; Otherwise, $T_{t-l}^{HT***} = T_{t-l}^{HT**}$ because $V_t = 0$. Please see $T_{t-l}^{HT***}$ in Figure C1C. The use of $T_{t-l}^{H***}$ will not adjust for effects of *HT^NHW* on HDs but $V_t$ will do.

It is possible that exposure to *EHT^HW* may have its health effect with a long induction period such that it would increase *R(Y)* on the next heatwave. $V_t$ may adjust for this effect, depending on *K* in $V_t$. Not to adjust for this effect, *j*-th component of $V_t$ can be set to be zero if *t-j* is a HD. We may not want to adjust for this but include such increased risks as lag effects of *HW*. Finally, to estimate lag effects of *HW*, Model 3 can be modified to

$$\log(E[Y_t]) = \alpha + \beta_3 HW_t + \sum_{j_L=0}^{J_L} \gamma_{L,j_L} T_{t-j_L}^{LT} + \sum_{j_H=0}^{K} \gamma_{H,j_H} T_{t-j_H}^{HT***} + \delta V_t + \sum_{p=1}^{P} \zeta_p FHW_{t-p}$$

$$+ confounders$$

We used this model and estimated $\exp(\beta_3 + \zeta_1)=RR$. In this model, from the perspective of estimation of the effect of $HT^{NHW}$ and $EHT^{HW}$, $\gamma_{H,j_H}$ for $j_H \geq 0$ represents the effect of $HT^{NHW}$ on $R(Y)$ of non-HDs, $\delta$ represents the effect of $HT^{NHW}$ on $R(Y)$ of HDs; $\beta_3$ represents the effect of $EHT^{HW}$ on $R(Y)$ of HDs, and $\zeta_p$ for $p \geq 1$ represents the effect of $EHT^{HW}$ on $R(Y)$ of non-HDs following a HD. $\gamma_{L,j_L}$ for $j_L \geq 0$ represents the effect of $LT$ on $R(Y)$ of HDs and non-HDs.

*C.3. Splines to adjust for confounding by temperature.*

Model 3 is a piecewise regression regarding *T*. For adjustment purposes, non-linear temperature-response association can also be modeled by using a spline function such as natural cubic spline (ns). Instead of *T*, we use $T^*$. If $HW_t=0$, $T^*= T - OT$ and if $HW_t=1$, $T^*=0$ such that

$$\log(E[Y_t]) = \alpha + \beta_3 HW_t + ns(T_t^*) + \cdots + ns(T_{t-K}^*) + V_t^{ns} + confounders$$

where $V_t^{ns}$ differs depending on day *t*. If day *t* is the first day of a heatwave,

$$V_t^{ns} = (ns(T_{t-1} - OT), ns(T_{t-2} - OT), \ldots, ns(T_{t-L} - OT)),$$

if day *t* is the second day of a heatwave,

$$V_t^{ns} = (0, ns(T_{t-2} - OT), \ldots, ns(T_{t-L} - OT)),$$

…, if day *t* is the *L-1*th day of a heatwave,

$$V_t^{ns} = (0, 0, \ldots, ns(T_{t-(L-1)} - OT), ns(T_{t-L} - OT)),$$

or if day *t* is the $L^{th}$ day of a heatwave, $V_t^{ns} = (0,0,\ldots,0,0)$. If $V_t^{ns}$ does not have much information in a dataset, the estimation of parameters for $V_t^{ns}$ may not be possible. $V_t$ can be used instead. Lag variables of $T^*$, $T_{t-K}^*$, can be generated in the two ways as described above.

To include $\zeta_x FHW_{t,x}$, $T_{t-l}^{**}$ or $T_{t-l}^{***}$ should be used instead. $T_{t-l}^{**}$ is the *l*-lag variable of $T_t^*$. Similarly,

$$\log(E[Y_t]) = \alpha + \beta_3 HW_t + ns(T_t^{**}) + \cdots + ns(T_{t-K}^{**}) + \zeta_1 FHW_{t-1} + \ldots + \zeta_X FHW_{t-X}$$
$$+ confounders$$

**Figure C1. An excerpt of daily time-series for temperature variables, a heatwave indicator, and an indicator of the subsequent day after heatwaves**

Note: The yellow-highlighted rows indicate heatwave days. For illustration, here, we defined heatwave as $T$ exceeding the 97th percentile (27.6°C) of the year-round $T$ distribution for at least two consecutive days. The data analysis in the main text used the 99th percentile (28.7°C), not the 97th percentile.

**A. Data-structure**

Heatwave indicator → $HW_t$, $FHW_{t-1}$

Basic temperature variables (piecewise regression) → $T_t$, $T_{t-1}$, $T_t^{LT}$, $T_{t-1}^{LT}$, $T_t^{HT}$, $T_{t-1}^{HT}$

(New variable introduced in this study) A variable to adjust for temperature as a confounder → $V_t$

| Year | Month | Day | Day of the Week | $HW_t$ | $FHW_{t-1}$ | $T_t$ | $T_{t-1}$ | $T_t^{LT}$ | $T_{t-1}^{LT}$ | $T_t^{HT}$ | $T_{t-1}^{HT}$ | $T_t^{HT*}$ | $T_{t-1}^{HT*}$ | $T_{t-1}^{HT**}$ | $T_{t-1}^{HT***}$ | $V_t$ | $T_t^*$ | $T_{t-1}^*$ | $T_{t-1}^{**}$ | $T_{t-1}^{***}$ |
|---|---|---|---|---|---|---|---|---|---|---|---|---|---|---|---|---|---|---|---|---|
| 2006 | 7 | 27 | Thursday | 0 | 0 | 21.0 | 22.5 | -0.8 | 0 | 0 | 0.8 | 0 | 0.8 | 0.8 | 0.8 | 0 | -0.8 | 0.8 | 0.8 | 0.8 |
| 2006 | 7 | 28 | Friday | 0 | 0 | 21.7 | 21.0 | -0.1 | -0.8 | 0 | 0 | 0 | 0 | 0 | 0 | 0 | -0.1 | -0.8 | -0.8 | -0.8 |
| 2006 | 7 | 29 | Saturday | 0 | 0 | 24.3 | 21.7 | 0 | -0.1 | 2.5 | 0 | 2.5 | 0 | 0 | 0 | 0 | 2.5 | -0.1 | -0.1 | -0.1 |
| 2006 | 7 | 30 | Sunday | 0 | 0 | 26.0 | 24.3 | 0 | 0 | 4.3 | 2.5 | 4.3 | 2.5 | 2.5 | 2.5 | 0 | 4.3 | 2.5 | 2.5 | 2.5 |
| 2006 | 7 | 31 | Monday | 0 | 0 | 26.5 | 26.0 | 0 | 0 | 4.7 | 4.3 | 4.7 | 4.3 | 4.3 | 4.3 | 0 | 4.7 | 4.3 | 4.3 | 4.3 |
| 2006 | 8 | 1 | Tuesday | 0 | 0 | 27.6 | 26.5 | 0 | 0 | 5.8 | 4.7 | 5.8 | 4.7 | 4.7 | 4.7 | 0 | 5.8 | 4.7 | 4.7 | 4.7 |
| 2006 | 8 | 2 | Wednesday | 1 | 0 | 28.0 | 27.6 | 0 | 0 | 6.2 | 5.8 | 0 | 0 | 5.8 | 0 | 5.8 | 0 | 0 | 5.8 | 0 |
| 2006 | 8 | 3 | Thursday | 1 | 0 | 28.4 | 28.0 | 0 | 0 | 6.7 | 6.2 | 0 | 0 | 0 | 0 | 0 | 0 | 0 | 0 | 0 |
| 2006 | 8 | 4 | Friday | 1 | 0 | 29.5 | 28.4 | 0 | 0 | 7.8 | 6.7 | 0 | 0 | 0 | 0 | 0 | 0 | 0 | 0 | 0 |
| 2006 | 8 | 5 | Saturday | 1 | 0 | 29.1 | 29.5 | 0 | 0 | 7.3 | 7.8 | 0 | 0 | 0 | 0 | 0 | 0 | 0 | 0 | 0 |
| 2006 | 8 | 6 | Sunday | 1 | 0 | 28.4 | 29.1 | 0 | 0 | 6.7 | 7.3 | 0 | 0 | 0 | 0 | 0 | 0 | 0 | 0 | 0 |
| 2006 | 8 | 7 | Monday | 0 | 1 | 27.2 | 28.4 | 0 | 0 | 5.5 | 6.7 | 5.5 | 6.7 | 0 | 0 | 0 | 5.5 | 6.7 | 0 | 0 |
| 2006 | 8 | 8 | Tuesday | 1 | 0 | 28.7 | 27.2 | 0 | 0 | 7.0 | 5.5 | 0 | 0 | 5.5 | 0 | 5.5 | 0 | 0 | 5.5 | 0 |
| 2006 | 8 | 9 | Wednesday | 1 | 0 | 29.3 | 28.7 | 0 | 0 | 7.5 | 7.0 | 0 | 0 | 0 | 0 | 0 | 0 | 0 | 0 | 0 |
| 2006 | 8 | 10 | Thursday | 1 | 0 | 28.7 | 29.3 | 0 | 0 | 6.9 | 7.5 | 0 | 0 | 0 | 0 | 0 | 0 | 0 | 0 | 0 |
| 2006 | 8 | 11 | Friday | 0 | 1 | 26.8 | 28.7 | 0 | 0 | 5.0 | 6.9 | 5.0 | 6.9 | 0 | 0 | 0 | 5.0 | 6.9 | 0 | 0 |
| 2006 | 8 | 12 | Saturday | 0 | 0 | 27.3 | 26.8 | 0 | 0 | 5.6 | 5.0 | 5.6 | 5.0 | 5.0 | 5.0 | 0 | 5.6 | 5.0 | 5.0 | 5.0 |

(New variable introduced in this study) An indicator variable to estimate one-day lag effect of $HW$

Basic temperature variables

(New variable introduced in this study) Variables to adjust for temperature as a confounder (piecewise regression)
Note. $T_{t-1}^{H***}= T_{t-1}^{H**} - V_t$ to remove redundancy

(New variable introduced in this study) Variables to adjust for temperature as a confounder (splines)
Note. $T_{t-1}^{***}= T_{t-1}^{**} - V_t$ to remove redundancy

## B. Highlights for adjustment for confounding by temperature

| Year | Month | Day | Day of the Week | $HW_t$ | $FHW_{t-1}$ | $T_t$ | $T_{t-1}$ | $T_t^{LT}$ | $T_{t-1}^{LT}$ | $T_t^{HT}$ | $T_{t-1}^{HT}$ | $T_t^{HT*}$ | $T_{t-1}^{HT*}$ | $T_{t-1}^{HT**}$ | $T_{t-1}^{HT***}$ | $V_t$ | $T_t^*$ | $T_{t-1}^*$ | $T_{t-1}^{**}$ | $T_{t-1}^{***}$ |
|---|---|---|---|---|---|---|---|---|---|---|---|---|---|---|---|---|---|---|---|---|
| 2006 | 7 | 27 | Thursday | 0 | 0 | 21.0 | 22.5 | -0.8 | 0 | 0 | 0.8 | 0 | 0.8 | 0.8 | 0.8 | 0 | -0.8 | 0.8 | 0.8 | 0.8 |
| 2006 | 7 | 28 | Friday | 0 | 0 | 21.7 | 21.0 | -0.1 | -0.8 | 0 | 0 | 0 | 0 | 0 | 0 | 0 | -0.1 | -0.8 | -0.8 | -0.8 |
| 2006 | 7 | 29 | Saturday | 0 | 0 | 24.3 | 21.7 | 0 | -0.1 | 2.5 | 0 | 2.5 | 0 | 0 | 0 | 0 | 2.5 | -0.1 | -0.1 | -0.1 |
| 2006 | 7 | 30 | Sunday | 0 | 0 | 26.0 | 24.3 | 0 | 0 | 4.3 | 2.5 | 4.3 | 2.5 | 2.5 | 2.5 | 0 | 4.3 | 2.5 | 2.5 | 2.5 |
| 2006 | 7 | 31 | Monday | 0 | 0 | 26.5 | 26.0 | 0 | 0 | 4.7 | 4.3 | 4.7 | 4.3 | 4.3 | 4.3 | 0 | 4.7 | 4.3 | 4.3 | 4.3 |
| 2006 | 8 | 1 | Tuesday | 0 | 0 | 27.6 | 26.5 | 0 | 0 | 5.8 | 4.7 | 5.8 | 4.7 | 4.7 | 4.7 | 0 | 5.8 | 4.7 | 4.7 | 4.7 |
| 2006 | 8 | 2 | Wednesday | 1 | 0 | 28.0 | 27.6 | 0 | 0 | 6.2 | 5.8 | 0 | 0 | 5.8 | 0 | 5.8 | 0 | 0 | 5.8 | 0 |
| 2006 | 8 | 3 | Thursday | 1 | 0 | 28.4 | 28.0 | 0 | 0 | 6.7 | 6.2 | 0 | 0 | 0 | 0 | 0 | 0 | 0 | 0 | 0 |
| 2006 | 8 | 4 | Friday | 1 | 0 | 29.5 | 28.4 | 0 | 0 | 7.8 | 6.7 | 0 | 0 | 0 | 0 | 0 | 0 | 0 | 0 | 0 |
| 2006 | 8 | 5 | Saturday | 1 | 0 | 29.1 | 29.5 | 0 | 0 | 7.3 | 7.8 | 0 | 0 | 0 | 0 | 0 | 0 | 0 | 0 | 0 |
| 2006 | 8 | 6 | Sunday | 1 | 0 | 28.4 | 29.1 | 0 | 0 | 6.7 | 7.3 | 0 | 0 | 0 | 0 | 0 | 0 | 0 | 0 | 0 |
| 2006 | 8 | 7 | Monday | 0 | 1 | 27.2 | 28.4 | 0 | 0 | 5.5 | 6.7 | 5.5 | 6.7 | 0 | 0 | 0 | 5.5 | 6.7 | 0 | 0 |
| 2006 | 8 | 8 | Tuesday | 1 | 0 | 28.7 | 27.2 | 0 | 0 | 7.0 | 5.5 | 0 | 0 | 5.5 | 0 | 5.5 | 0 | 0 | 5.5 | 0 |
| 2006 | 8 | 9 | Wednesday | 1 | 0 | 29.3 | 28.7 | 0 | 0 | 7.5 | 7.0 | 0 | 0 | 0 | 0 | 0 | 0 | 0 | 0 | 0 |
| 2006 | 8 | 10 | Thursday | 1 | 0 | 28.7 | 29.3 | 0 | 0 | 6.9 | 7.5 | 0 | 0 | 0 | 0 | 0 | 0 | 0 | 0 | 0 |
| 2006 | 8 | 11 | Friday | 0 | 1 | 26.8 | 28.7 | 0 | 0 | 5.0 | 6.9 | 5.0 | 6.9 | 0 | 0 | 0 | 5.0 | 6.9 | 0 | 0 |
| 2006 | 8 | 12 | Saturday | 0 | 0 | 27.3 | 26.8 | 0 | 0 | 5.6 | 5.0 | 5.6 | 5.0 | 5.0 | 5.0 | 0 | 5.6 | 5.0 | 5.0 | 5.0 |

C. Highlights for adjustment for or estimation of lag effects of heatwaves.

| Year | Month | Day | Day of the Week | $HW_t$ | $FHW_{t-1}$ | $T_t$ | $T_{t-1}$ | $T_t^{LT}$ | $T_{t-1}^{LT}$ | $T_t^{HT}$ | $T_{t-1}^{HT}$ | $T_t^{HT*}$ | $T_{t-1}^{HT*}$ | $T_{t-1}^{HT**}$ | $T_{t-1}^{HT***}$ | $V_t$ | $T_t^*$ | $T_{t-1}^*$ | $T_{t-1}^{**}$ | $T_{t-1}^{***}$ |
|---|---|---|---|---|---|---|---|---|---|---|---|---|---|---|---|---|---|---|---|---|
| 2006 | 7 | 27 | Thursday | 0 | 0 | 21.0 | 22.5 | -0.8 | 0 | 0 | 0.8 | 0 | 0.8 | 0.8 | 0.8 | 0 | -0.8 | 0.8 | 0.8 | 0.8 |
| 2006 | 7 | 28 | Friday | 0 | 0 | 21.7 | 21.0 | -0.1 | -0.8 | 0 | 0 | 0 | 0 | 0 | 0 | 0 | -0.1 | -0.8 | -0.8 | -0.8 |
| 2006 | 7 | 29 | Saturday | 0 | 0 | 24.3 | 21.7 | 0 | -0.1 | 2.5 | 0 | 2.5 | 0 | 0 | 0 | 0 | 2.5 | -0.1 | -0.1 | -0.1 |
| 2006 | 7 | 30 | Sunday | 0 | 0 | 26.0 | 24.3 | 0 | 0 | 4.3 | 2.5 | 4.3 | 2.5 | 2.5 | 2.5 | 0 | 4.3 | 2.5 | 2.5 | 2.5 |
| 2006 | 7 | 31 | Monday | 0 | 0 | 26.5 | 26.0 | 0 | 0 | 4.7 | 4.3 | 4.7 | 4.3 | 4.3 | 4.3 | 0 | 4.7 | 4.3 | 4.3 | 4.3 |
| 2006 | 8 | 1 | Tuesday | 0 | 0 | 27.6 | 26.5 | 0 | 0 | 5.8 | 4.7 | 5.8 | 4.7 | 4.7 | 4.7 | 0 | 5.8 | 4.7 | 4.7 | 4.7 |
| 2006 | 8 | 2 | Wednesday | 1 | 0 | 28.0 | 27.6 | 0 | 0 | 6.2 | 5.8 | 0 | 0 | 5.8 | 0 | 5.8 | 0 | 0 | 5.8 | 0 |
| 2006 | 8 | 3 | Thursday | 1 | 0 | 28.4 | 28.0 | 0 | 0 | 6.7 | 6.2 | 0 | 0 | 0 | 0 | 0 | 0 | 0 | 0 | 0 |
| 2006 | 8 | 4 | Friday | 1 | 0 | 29.5 | 28.4 | 0 | 0 | 7.8 | 6.7 | 0 | 0 | 0 | 0 | 0 | 0 | 0 | 0 | 0 |
| 2006 | 8 | 5 | Saturday | 1 | 0 | 29.1 | 29.5 | 0 | 0 | 7.3 | 7.8 | 0 | 0 | 0 | 0 | 0 | 0 | 0 | 0 | 0 |
| 2006 | 8 | 6 | Sunday | 1 | 0 | 28.4 | 29.1 | 0 | 0 | 6.7 | 7.3 | 0 | 0 | 0 | 0 | 0 | 0 | 0 | 0 | 0 |
| 2006 | 8 | 7 | Monday | 0 | 1 | 27.2 | 28.4 | 0 | 0 | 5.5 | 6.7 | 5.5 | 6.7 | 0 | 0 | 0 | 5.5 | 6.7 | 0 | 0 |
| 2006 | 8 | 8 | Tuesday | 1 | 0 | 28.7 | 27.2 | 0 | 0 | 7.0 | 5.5 | 0 | 0 | 5.5 | 0 | 5.5 | 0 | 0 | 5.5 | 0 |
| 2006 | 8 | 9 | Wednesday | 1 | 0 | 29.3 | 28.7 | 0 | 0 | 7.5 | 7.0 | 0 | 0 | 0 | 0 | 0 | 0 | 0 | 0 | 0 |
| 2006 | 8 | 10 | Thursday | 1 | 0 | 28.7 | 29.3 | 0 | 0 | 6.9 | 7.5 | 0 | 0 | 0 | 0 | 0 | 0 | 0 | 0 | 0 |
| 2006 | 8 | 11 | Friday | 0 | 1 | 26.8 | 28.7 | 0 | 0 | 5.0 | 6.9 | 5.0 | 6.9 | 0 | 0 | 0 | 5.0 | 6.9 | 0 | 0 |
| 2006 | 8 | 12 | Saturday | 0 | 0 | 27.3 | 26.8 | 0 | 0 | 5.6 | 5.0 | 5.6 | 5.0 | 5.0 | 5.0 | 0 | 5.6 | 5.0 | 5.0 | 5.0 |

**Appendix D. Regarding exposure misclassification.**

Many epidemiological studies use *T* measured from monitoring stations or modeled estimates of *T* at residence (10, 11), not *E*, and although they are correlated, not perfectly so. Some may argue that such use of *T* introduces exposure misclassification because individuals are not necessarily exposed to a heatwave (i.e., through $P_1$ and $P_2$ in Figure 1). This viewpoint holds if *E-Y* relation is the causal effect of interest. However, there are other viewpoints that advocate the use of *HW* and *T*. First, *HW-Y* relation can be of interest, which is policy-relevant as we noted in the main text. *HW-Y* relation would represent a population averaged total effect that presents an overall picture of the health impact in a population of interest, including exposed and unexposed groups. We call this crude *HW-Y* relation 'with respect to *E*' because this would be a weighted average of two stratum-specific relations: *HW-Y* relation for *E*=1 (those who would be exposed to heatwaves during heatwaves) and *HW-Y* relation for *E*=0 (those who would be unexposed to heatwaves during heatwaves). If a heatwave is harmful, *HW-Y* relation for *E*=1 will be positive and *HW-Y* relation for *E*=0 will be the null. Crude *HW-Y* relation would be meaningful to estimate the health impact of the total effect of *HW* (e.g., attributable fractions/numbers) when the size of the population is (usually) known but the size of exposed and unexposed groups is unknown. If *HW-Y* relation is of interest, the use of *HW*, not *E*, will not necessarily result in exposure misclassification, demonstrating that the use of *HW* or *E* should depend on what to infer.